\documentclass[%
preprint,
superscriptaddress,
 amsmath,amssymb,
 aps,
]{revtex4-2}

\usepackage{graphicx}
\usepackage{dcolumn}
\usepackage{bm}
\usepackage[T1]{fontenc}
\usepackage[utf8]{inputenc}
\usepackage{textcomp}

\begin{document}

\title{Transport of topologically protected photonic waveguide on chip}

\author{
    Sai Yan $^1$$^,$$^2$ , Jingnan Yang $^3$ , Shushu Shi  $^1$$^,$$^2$ , Zhanchun Zuo  $^1$ , Can Wang  $^1$$^,$$^4$$^,$$^5$ , Xiulai Xu  $^{3\ast}$\\
    $^1$ Beijing National Laboratory for Condensed Matter Physics, Institude of Physics, Chinese Academy of Sciences, Beijing 100190, China\\
    $^2$ CAS Center for Excellence in Topological Quantum Computation and School of Physical Sciences, University of Chinese Academy of Sciences, Beijing 100049, China\\
    $^3$ State Key Laboratory for Mesoscopic Physics and Frontiers Science Center for Nano-optoelectronics, School of Physics, Peking University, 100871 Beijing, China\\
    $^4$ Songshan Lake Materials Laboratory, Dongguan, Guangdong 523808, China\\
    $^5$ e-mail: canwang@iphy.ac.cn\\
    $^\ast$ xlxu@pku.edu.cn
}

\begin{abstract}

We propose a new design on integrated optical devices on-chip with an extra width degree of freedom by using a photonic crystal waveguide with Dirac points between two photonic crystals with opposite valley Chern numbers. With such an extra waveguide, we demonstrate numerically that the topologically protected photonic waveguide keeps properties of valley-locking and immunity to defects. Due to the design flexibility of the width-tunable topologically protected photonic waveguide,  many unique on-chip integrated devices have been proposed, such as energy concentrators with a concentration efficiency improvement by more than one order of magnitude, topological photonic power splitter with arbitrary power splitting ratio. The topologically protected photonic waveguide with the width degree of freedom could be beneficial for scaling up photonic devices, which provides a new flexible platform to implement integrated photonic networks on chip.

\end{abstract}

\maketitle

\section{Introduction}
Due to the advantages of low energy cost and broad bandwidth in processing information, on-chip nanophotonic devices have been widely studied in recent years \cite{sun2015single,kok2007linear,xiao2021position,shu2022chiral}, which become an important platform for constructing all-optical connectivity, all-optical computing, and all-optical networks \cite{sun2015single,kok2007linear,ji2016prospects,caulfield2010future,xiao2021chiral}. However, imperfect fabrication of nanophotonic devices is often unavoidable,  which lowers the performance of the devices \cite{xie2021optimization}. The emergence of topological photonics has provided a way to solve the functional influences of fabrication errors of nanophotonic devices \cite{xie2021optimization,ma2016all,gao2017valley,wu2017direct}, which shows great potential to control the propagation of light.

In particular, the topological edge states which exist at the interface between two valley photonic crystals (PhCs) with opposite valley Chern numbers or band inversion exhibit novel properties, such as robustness against defects and unidirectional transmission \cite{ma2016all,gao2017valley,wu2017direct,gao2018topologically,noh2018observation,chen2018valley,shalaev2019robust,he2019silicon,chen2019valley,zhang2019valley,yang2020terahertz,kumar2022terahertz,kumar2022topological-2}. So far, a large number of on-chip topological photonic devices exhibiting robustness have been demonstrated, such as topological lasers \cite{gong2020topological,liu2022topological,zhang2020low,zhong2020topological},  topological optical switches \cite{riza1999reconfigurable,qi2021all,kumar2022active}, topologically protected photonic waveguides \cite{shalaev2019robust,mehrabad2020chiral,haldane2008possible,wang2009observation,fang2012realizing,xi2020topological,kumar2022phototunable}, topological filters \cite{gu2021topological,chen2019direct,lu2019spin,yang2020topological}, and topological sensors\cite{kumar2022topological}. Most of these topological nanophotonic devices show a better performance. For example, the topologically protected photonic waveguides can have a larger operating bandwidth, lower propagation losses, a smaller footprint and a higher operation efficiency at telecommunication wavelengths \cite{xie2021topological,gao2017valley,gao2018topologically,shalaev2019robust,he2019silicon,zhang2019valley,liu2022terahertz}. However, since topological edge states are usually localized at the interface between the valley PhCs \cite{ma2016all,gao2017valley,wu2017direct,gao2018topologically,noh2018observation,chen2018valley,shalaev2019robust,he2019silicon,chen2019valley,zhang2019valley,yang2020terahertz,xie2021topological}, the width of the topologically protected photonic waveguides is not tunable. A tunable width of waveguides provides not only a degree of freedom (DOF) in designing photonic devices\cite{xiao2021chiral,xiao2021position,wang2021topological,chen2021photonic}, but also an opportunity to interface with other photonic devices with a high flexibility. Therefore, it is highly desired to introduce the width DOF in topologically protected photonic waveguides. The introduction of width DOF has been proposed and experimentally demonstrated for devices working microwave regime \cite{wang2021topological,chen2021photonic}. To implement on-chip integrated optical networks \cite{nagarajan2005large,koch1991semiconductor,son2018high,komljenovic2016heterogeneous}, here we expand it to the near infrared optical wavelength range.

In this paper, we propose to introduce width DOF into a topologically protected photonic waveguide (TPPW) to more flexibly manipulate the light transport. PhCs with Dirac points are sandwiched by two PhCs with opposite valley-Chern numbers, which constitutes the TPPW with a high flexibility. The TPPW with the introduction of width DOF keeps the characteristics of gapless dispersion, valley-momentum locking and robustness to defects. Taking advantage of these properties, we design a multi-level energy concentrator by abruptly reducing the waveguide width. Energy intensity is enhanced by from 2.48 to 23.38 times by putting a photonic structure with bandgap as a mirror at the end of the waveguide. The photonic energy concentrator could be used for applications in energy convergences \cite{piao2021ultra} and nonlinear optics \cite{lan2020nonlinear,xie2020cavity}. In addition, a topological photonic power splitter with arbitrary coupling ratios is also proposed by introducing width DOF, which could be useful for scaling up photonic devices on chip with low losses.

\section{Structure Design}

The topological valley PhC here contains three domains A, B, and C (Fig. 1(a)), which are composed of a honeycomb lattice with two different sizes of triangular airholes, as shown in the insets of Fig. 1(a). The side lengths of the triangular airholes are described by L1 and L2, respectively. Dirac degeneracy exists in $K$ valley of the PhCs in the B domain when the two triangular airholes are the same size (L1=L2=82 nm). When L1 $\neq$ L2, the spatial inversion of the PhCs in the A and C domains is broken with the formation of a bandgap (as shown in Fig. 1(b)). The PhCs in the A or C domains show nonzero Berry curvatures, with opposite signs at $K$ and $K'$ valleys. Due to the time-reversal symmetry, the Berry curvature is equal to zero by integrating over the entire Brillouin zone (BZ). It’s worth noting that the valley Chern number is nonzero since we only need to integrate the Berry curvature over the half of the BZ around $K$/$K'$ valley \cite{he2019silicon,shalaev2019robust}. The inset of Fig. 1(a) shows the rhomboid units of three topologically distinct PhCs. The larger triangular airholes in the PhC of the A domain (blue) and C domain (red) are oppositely orientated. The sign of the Berry curvature at each valley will be reversed by swapping the sizes of the two triangular airholes. Therefore, the PhCs in the A and C domains appear with opposite valley Chern numbers and different valley phases. To illustrate the phase transition visually, we simulated the z-component of the magnetic field (H$_z$) phase profiles of the PhCs in the A and C domains at $K$ point. Figure 1(c) and 1(f) show valley phase locking with a clockwise vortex at $K$ point in the upper band (321.11 THz) of the PhCs in the A domain and the lower band (293.14 THz) of the PhCs in the C domain, respectively. Figure 1(d) and 1(e) show another valley phase locking with an anticlockwise vortex at $K$ point in the lower band of the PhCs in the A domain and the upper band of the PhCs in the C domain, respectively. The red and blue arrows point to the phase profile increasing by 2$\pi$ phase either anticlockwise or clockwise around the center of the unit cell. Due to the time-reversal symmetry, the counterparts have opposite vertexes at $K'$ point. The topological edge states exist the interface between two topologically distinct PhCs with the opposite optical vortexes \cite{ma2016all,gao2017valley,wu2017direct,gao2018topologically,noh2018observation,chen2018valley,shalaev2019robust,he2019silicon,chen2019valley,zhang2019valley,yang2020terahertz,xie2021topological}. By introducing a PhC in the B domain into the interface, the topological edge states still exist.

\begin{figure}[h!]
	\centering\includegraphics[width=13cm]{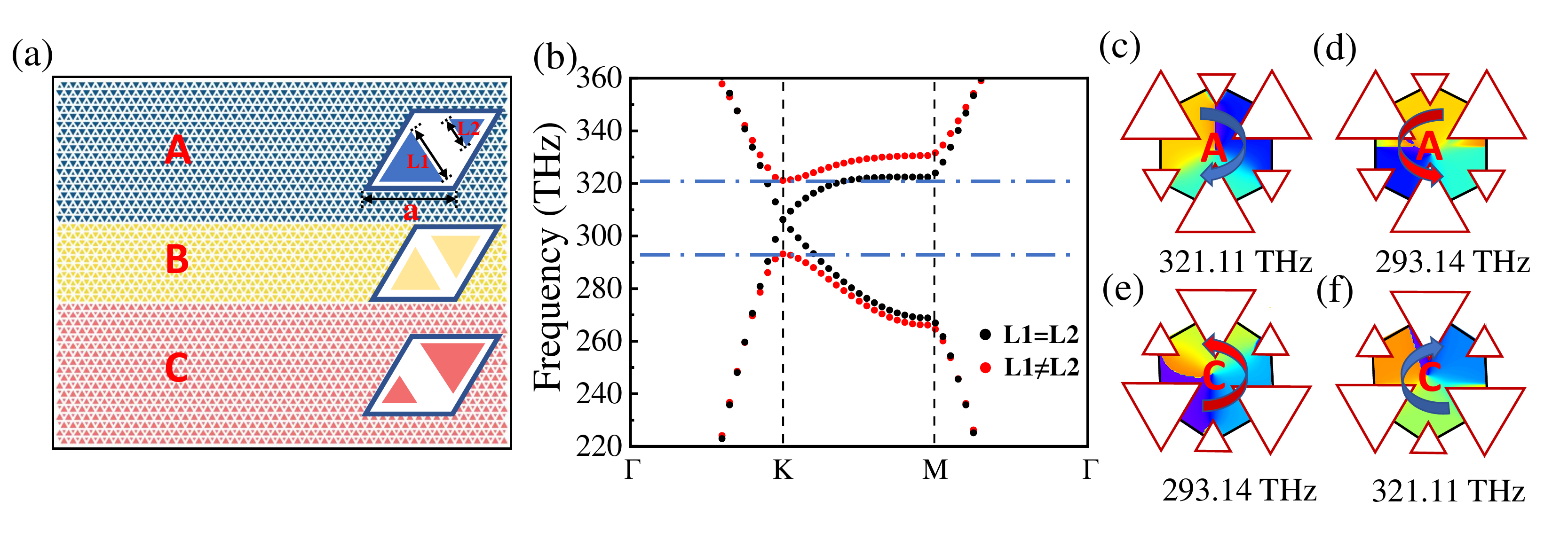}
	\caption{(a) Schematic diagram of the topological protected photonic waveguide (TPPW). The TPPW consists of three domains, A$\mid$$B_x$$\mid$C, where the unit cells of PhCs in A and C domains have broken-inversion symmetry and that in B domain has inversion symmetry. The structure parameters are L1=196 nm and L2=84 nm for A domain (L1=L2=151.2 nm for B domain, L1=84 nm and L2=196 nm for C domain), the lattice constant a=280 nm, and the thickness of the structure d=150 nm. (b) Band diagrams of PhC in B domain (L1=L2) with Dirac points, and band diagrams of PhC on A or C domain (L1$\neq$L2) with a bandgap. (c)-(f) $H_z$ phase profiles of PhC in A domain and PhC in C domain at $K$ point in the upper (321.11 THz) and lower (293.14 THz) bands.}
\end{figure}

Figure 2(a) depicts the projected band structure of the TPPW with layers $x$=5 of B domain ($A\mid$$B_5$$\mid$$C$). The red line inside the bandgap shows a gapless band. With a pair of opposite group velocities connected to $K$ and $K'$, the gapless band has the same valley-locked property as the edge state in the A$\mid$C domain interface. This state is referred to as the topological valley-locked waveguide state (TVWS). In addition, there are higher-order non-topological waveguide states (NTWS) with gapped dispersion, represented by the symbols $0^+$th, $0^-$th, $1^+$st, $1^-$st (as shown in Fig. 2(a)). It is worth noting that different from TVWSs, NTWSs are gapped and lack the valley-locked property. The gap between the $0^+$th band and the $0^-$th band is called the topological frequency window (light grey area in Fig. 2(a)), where only the TVWS exists. From Fig. 2(b), we can see that the width of the topological frequency window varies with the number of layers $x$ in B domain. As the number of layer $x$ in B domain increases, the width of the topological frequency window narrows with higher-order NTWSs such as $1^+$st, $1^-$st, $2^+$nd and $2^-$nd. In the extreme case when the layer $x$ in the B domain is large enough, the topological frequency window eventually closes and the bulk-edge correspondence between A and C domains gradually weakens. As a result, the waveguide band structure becomes the bulk band structure of the B domain.

To explain the origin of the topological guided mode in the TVWSs, we turn our attention to the $k$·$p$ perturbation method \cite{notomi2000theory,wu2017direct,mei2012first}.
The effective Hamiltonian around the $K$ valley which represent the PhCs in the A, B and C domains can be written as.
\begin{eqnarray}
	\delta H_K(\delta k) = c_D\delta k_x\sigma_x + c_D\delta k_y\sigma_y + mc_D^{2}\sigma_z	
\end{eqnarray}
where $m$ is the effective mass with $m$$\textless$0 for A crystal, $m$=0 for B crystal, and $m$$\textgreater$0 for C crystal, $c_D$ is the group velocity or the slope of the linear Dirac cone in B crystal, $\delta \boldsymbol{k}$ = $\boldsymbol{k}$ - $\boldsymbol{k_K}$ is the displacement of wave vector $\boldsymbol{k}$ to the K valley and $\sigma_i$ (i=x, y, z) are Pauli matrices. It is obvious that the effective Hamiltonian around the $K'$ valley can be obtained by applying a time reversal operation on the Hamiltonian at $K$ valley. From the eigenvalue equation	$\delta H_K \phi = \delta \omega \phi$,
we can obtain the dispersion relation,
\begin{eqnarray}
	\delta ^{2} \omega = c_D^{2}(\delta ^{2}k_x + \delta ^{2}k_y) + m^{2}c_D^{4}
\end{eqnarray}
which describes the dispersion relation of the A, B, and C crystals. Additionally, the topological guided mode $\phi_A$$_B$$_C$ in the TPPW exponentially attenuates along the $+Y$ direction in A domain and along the $-Y$ in C domain \cite{wang2020valley,chen2021photonic}. A specific solution with $\delta \omega = c_D\delta k_x$ can be obtained, where the slope $c_D$ is the identical to the bulk states in B domain ($\delta \omega = \pm c_D\delta k$). Therefore, we can conclude that the topological guided mode is a combination of the valley edge state at the A$\mid$C domain interface and the bulk states in B domain \cite{wang2020valley,chen2021photonic,wang2021topological,wang2022extended,chen2022robust}.

The frequency window of the photonic crystal waveguide directly affects the performance of the photonic device, where a wider frequency window will improve the operating bandwidth of the photonic devices. We compare the width of topological frequency window of valley-locked waveguide consisting of honeycomb lattices with two triangular and two circle airholes, respectively (as shown in Fig. 2(c)). In the topological valley photonic crystal with circle airholes, the radii of the circles airholes in the A/C domain are 66.7 nm and 42 nm, respectively. The radius of the circle airhole in the B domain is 54 nm. The lattice constant is 280 nm. We can see that a valley-locked waveguide with a honeycomb lattice with two triangular airholes has a much larger topological frequency window than one with a honeycomb lattice with two circle airholes, regardless of the number of layers in B domain. This is due to the difference in the band structures of the PhCs consisting of a honeycomb lattice with two triangle airholes and the PhCs consisting of a honeycomb lattice with two circle airholes. The former is a direct bandgap while the latter is an indirect bandgap, as shown in the inset in Fig. 2(c). The indirect bandgap of honeycomb lattice with two circle airholes might due to the smaller effective refraction index than that with two triangle airholes, which also modifies the dispersion in the gap. From practical point of view, TPPWs consisting of the PhCs with a direct bandgap has a larger topological frequency window than one with an indirect bandgap, which is important particually for devices with self-assembled quantum dots with a large size distribution \cite{xie2021topological}. Therefore, we choose PhCs consisting of a honeycomb lattice with two different sizes of triangular airholes to construct TPPWs in this work. In addition, the bandwidth of the devices is also related to the size of bandgap of the photonic crystals that make up the device and the waveguide interfaces that constitute the devices\cite{tan2022interfacial}. As the size of bandgap of the photonic crystal increases, i.e., L1/L2, the bandwidth of the corresponding topologically protected waveguide becomes larger. The maximum achievable bandwidth of the topologically protected waveguide is approximately 45 THz when experimental fabrication is considered. With all the above reasons considered, we choose the large triangle and small triangle side lengths of 196 nm, 84 nm, and zigzag interfaces.

How the topological guide modes transport in the TPPW with width DOF is very important for subsequent device design. In the most simple cases, we put red-marked point sources inside three waveguides with $x$ = 1, 5, 9, to excite the TVWSs, respectively. Figure 2(d) shows the H$_z$ field map of three TVWSs with $x$ = 1, 5, 9 at 302.08 THz. From three simulated H$_z$ field maps, we can get energy confined to the B domain to propagate between the A and C domains regardless of the number of layers in the B domain, which makes it possible to control the TVWSs transport by varying the TPPW width.
\section{Properties of the TVWSs}
\begin{figure}[h!]
	\centering\includegraphics[width=13cm]{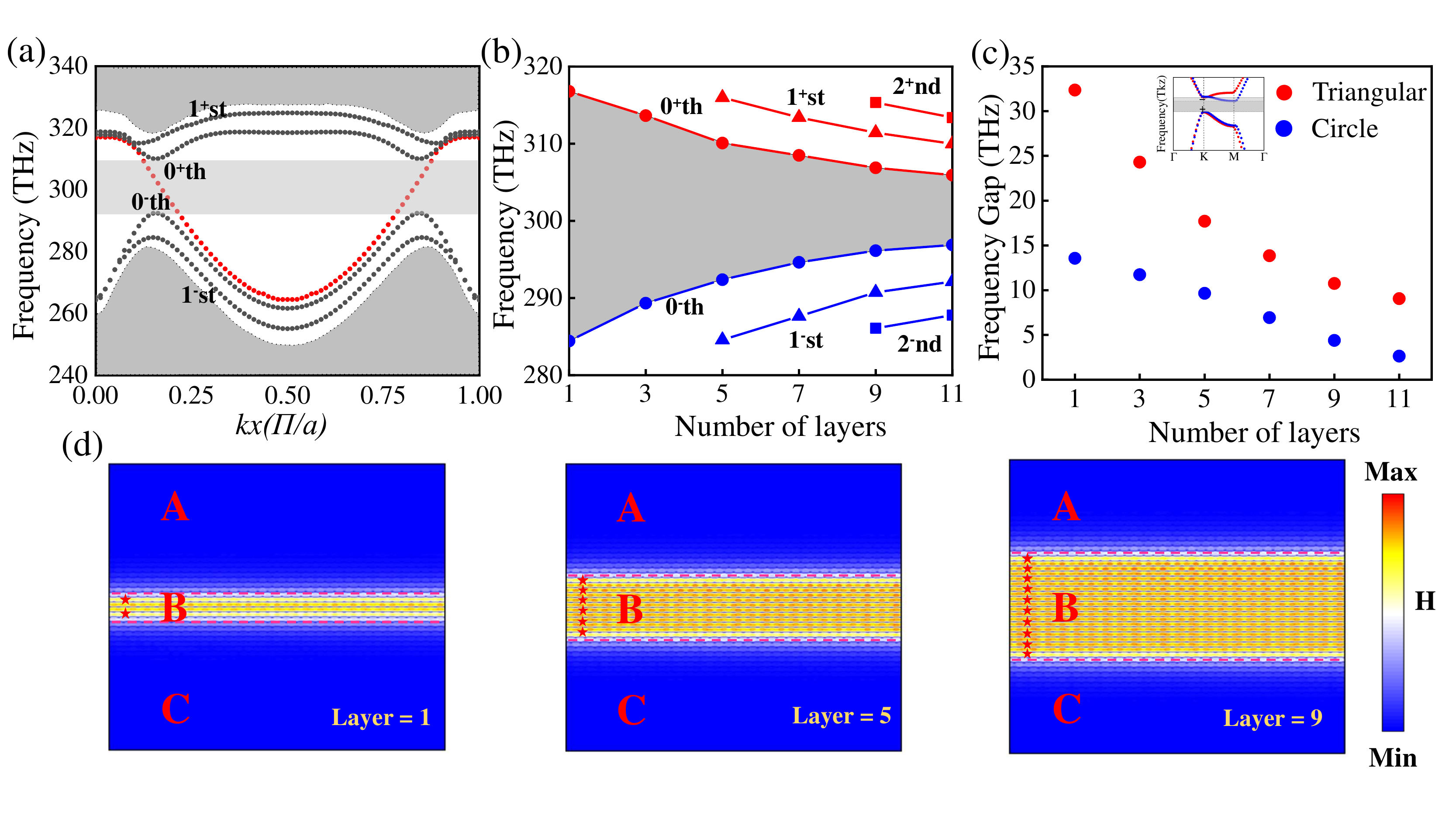}
	\caption{(a) Band diagram of $A\mid$$B_5$$\mid$$C$ waveguide. The red dotted lines show the topological edge state. The 0th and 1st bands marked with black lines, are other gapped waveguide states. The light grey area in the bulk gap of A or C domain is referred to the topological frequency window. (b) The width of the topological frequency window as a function of the number of layer $x$ in B domain. (c) The frequency gap of the topological frequency window of valley-locked waveguide consisting of honeycomb lattices with two sizes of triangular and circle airholes as a function of the number of layer $x$ in B domain. Inset: Band diagrams with different shapes. (d) H$_{z}$ field distributions of three straight TPPWs with $x$=1, 5, and 9.}
\end{figure}

To demonstrate the robustness against defects of the TVWSs, we introduce four distinct defects into B domain of the waveguides. The four structural defects are bulging, indentation, bending, and disorder, respectively. Figure 3(a)-3(d) display the simulated field distributions for the four waveguides with four types of structural defects at a frequency of 302.08 THz. We put the point source into input port 1 to excite the TVWSs and the monitor into the terminal of output port 4 to receive the signal. The TVWSs can transport forward while passing defects due to the protection of topology. We compare the energy intensity of the waveguide without any defects with the transmission fields measured at the positions labelled by the blue dots at the exits of the four waveguides in Fig. 3(a)-3(d). As shown in Fig. 3(e), regardless of the existence of defects, the TVWSs propagation within the topological frequency window (shaded region) keeps highly consistent with that of the topological frequency window without the defect. On the contrary, outside the shaded area, the simulated field intensity can become largely different due to the existence of other high-order edge states. The phase of TVWSs transmitted in the waveguide without defects is shown in Fig. 3(f), while the phase of TVWSs propagating in the waveguide shown in Fig. 3(a) after passing through the defects is shown in Fig. 3(g). We can see that the phases of two TVWSs are highly consistent. Figure 3(b)-3(d) are also consistent with the two phases diagrams above which implies that the propagation of TVWSs is immune to different defects \cite{wang2022extended}.

\begin{figure}[h!]
	\centering\includegraphics[width=13cm]{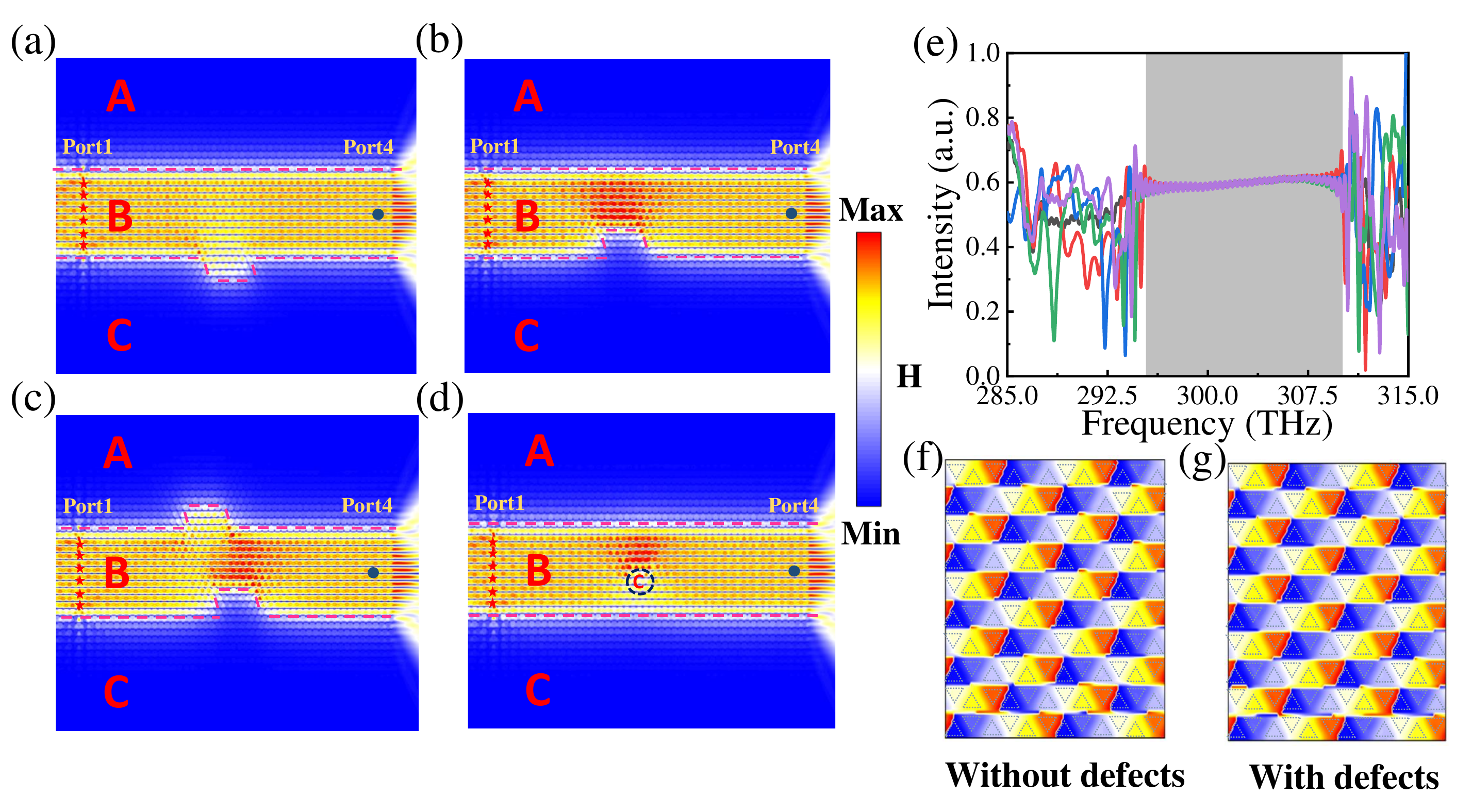}
	\caption{(a)-(d) Simulated photonic H$_{z}$ field distributions at 302.08 THz in the waveguides with four different defects: bulging, indentation, bending, and disorder. (e) The field intensities measured at the points represented by blue dots near the exits of the four waveguides with defects in Fig. 3(a)-(d). The black line is the field intensity of waveguide without defects. The great consistency of field intensities in the topological frequency windows indicated the immunity of the TVWSs to defects. The H$_{z}$ phase diagrams of the TVWSs after transmission without (f) and with defects (g).}
\end{figure}

To show the existence of the valley-locked effect, a valley-locked topological channel intersection is simulated based on the TVWSs. The device's configuration, as illustrated in Fig. 4(a) and 4(b), consists of five domains separated by the red dotted line, with the number of layers x=5 in B domain. We set ports 1-4 at the terminals of the B domain, placing point sources at port 1 as input ports to excite TVWSs, while ports 2-4 are considered as output ports to receive signals. The field distributions at 302.08 THz as shown in Fig. 4(a) and 4(b). Since TVWSs propagate in the A$\mid$$B_x$$\mid$C and C$\mid$$B_x$$\mid$A waveguides instead of the A$\mid$$B_x$$\mid$A and C$\mid$$B_x$$\mid$C waveguides \cite{wang2020valley}, TVWSs transmitted to output port 2 and port 3 are suppressed and the TVWSs locked to the K-valley propagate only along channel port 4, as in the case of Fig. 4(a). In contrast, TVWSs locked to the K-valley propagate along the channel toward port 2 and port 3 but not port 4 because TVWSs from the A$\mid$$B_x$$\mid$C and C$\mid$$B_x$$\mid$A waveguides are locked to the opposite valley, as shown in Fig. 4(b). To explain the splitting effect of the TVWSs more intuitively, we put the monitors on the light blue dashed lines as shown in Fig. 4(a) and 4(b) at ports 2, 3, and 4, respectively, to obtain the transmission spectra. Figure 4(c) shows that TVWSs mostly propagate along the channel toward port 4 in the topological frequency window with a transmittance of 98$\%$. Figure 4(d) shows that TVWSs barely propagate along the channel toward port 4 in the topological frequency window. It is worth noting that the transmittances of TVWSs through port 2 and port 3 are not the same due to the different types of waveguide interfaces. These simulation results indicate that TVWSs locked into different valleys only propagate in the corresponding channels with the extra width DOF.

\begin{figure}[h!]
	\centering\includegraphics[width=13cm]{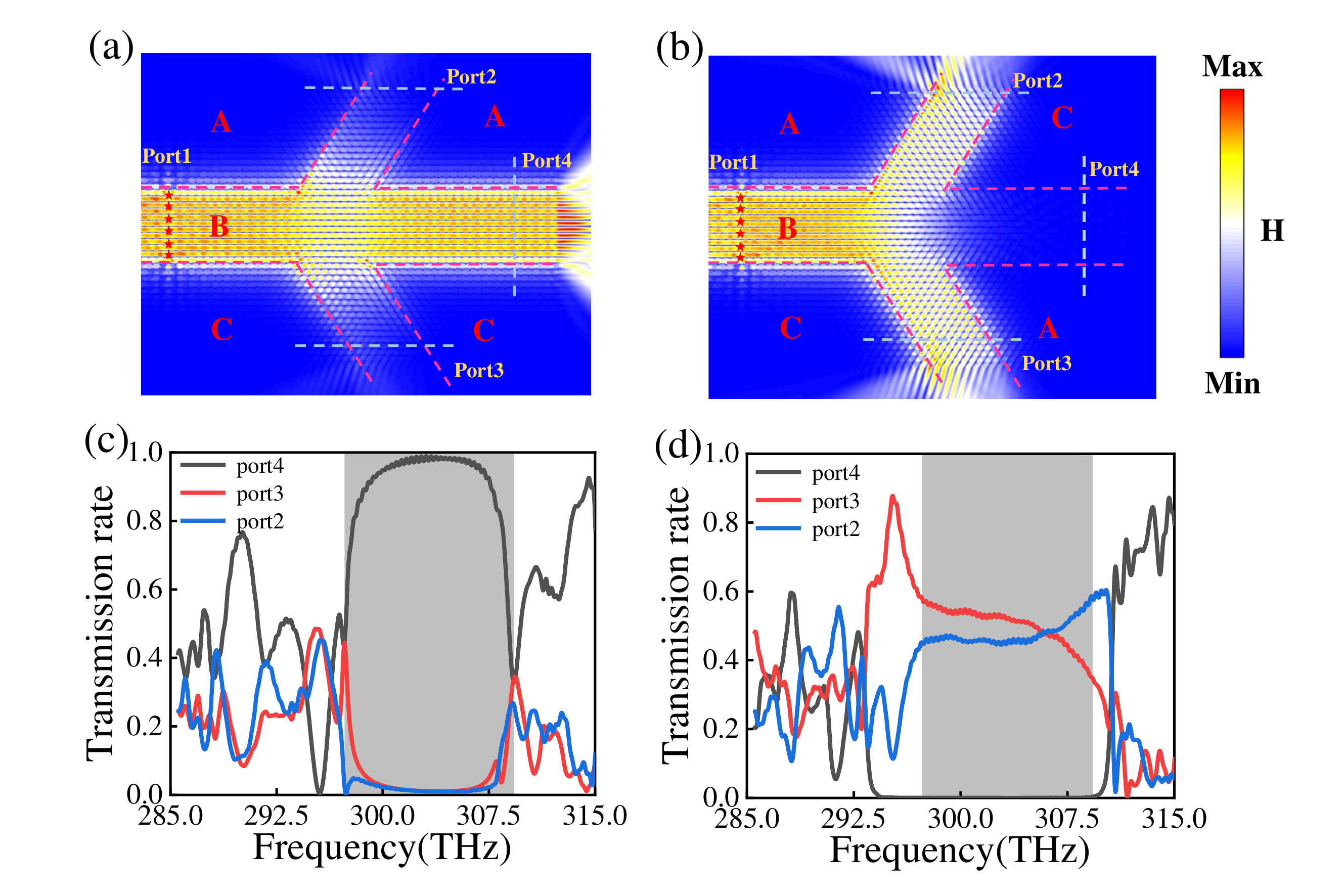}
	\caption{(a), (b) Simulated photonic H$_{z}$ field distributions under the excitation of point sources on the input port1 for two different configurations at 302.08 THz. Three ports labeled 2, 3, and 4 are located in B domain. (c), (d) The transmittance of TVWSs through port2, port3, and port4 for Fig. 4(a) and Fig. 4(b), respectively.}
\end{figure}
\section{Applications of the TVWSs}
To realize on-chip optical integration network on small footprint and high degree of integration, the high-efficient energy output of on-chip optical integration network is one of the remaining challenges. With the robustness and valley-locking properties of TVWSs discussed above, we design an energy concentrator for energy harvesting. Figure 5(a) shows the schematic diagram of the topological concentrator. The first half of the topological concentrator consists of a waveguide with a sharp decrease in the number of layers in the B domain from 9 to 0.5. The sharp turns in the system are fixed at 60$^\circ$  or 120$^\circ$. We place point sources marked as red stars to excite TVWSs. When TVWSs enter the narrow part of the B domain, backscattering hardly occurs due to the valley-locked property \cite{chen2021photonic,wang2020valley}. The energy converges to the narrow part of the B domain, called the first-order convergence. The waveguide with x=0.5 in the B domain is terminated by a mirror made up of a photonic crystal, which constitutes the latter half of the topological concentrator. The photonic crystal structure as a mirror in Figure 5(a) is composed of a periodic array of a square lattice with square airhole. The length of the square airhole is 204.8 nm, the lattice constant is 320 nm and the thickness of the structure is 150 nm. The bandgap of the mirror ranges from 283.6 THz to 323.9 THz, while the bandwidth of the device in Fig. 5(a) ranges from 296.129 THz to 306.881 THz. Therefore, when the TVWSs that are within the bandwidth propagate along the topologically protected waveguide to the end, TVWSs cannot propagate inside mirrors. When the TVWSs transport to the terminal of the waveguide, the energy is further gathered due to the existence of the mirror and the near-conservation of the valley \cite{ma2016all,gao2018topologically,li2020topology}. As shown in Fig. 5(b), a large amount of energy is concentrated in a very small cavity, as we called the second-order convergence. To quantitatively assess the ability of the energy convergence, the intensity profiles are shown by simulations in Fig. 5(c) along the black, red and blue dotted lines in Fig. 5(a). It can be seen that from the inset of Fig. 5(c) the energy intensity along the red line is 2.48 times higher than the energy intensity along the black line. It is worth noting that the energy intensity along the blue line is 23.38 times higher than that along the black line, as shown in Fig. 5(c). The energy concentrator without backscattering could be used in the integrated photonic circuit. Meanwhile, the energy concentrator is essentially a cavity with a very small mode volume, as shown in Fig. 5(b), which could be used for studying cavity quantum electrodynamics and low threshold lasers \cite{xie2020cavity, xie2021topological, qian2018two, qian2019enhanced,zhang2020low}.

Figure 5(d) shows a topological photonic power splitter. By introducing the B domain into the A$\mid$C boundary, it provides a new degree of freedom to the power splitter. We design a relatively simple topological photonic power splitter to illustrate how the new degree of freedom comes into play. As shown in Fig. 5(d), the structural design of the power splitter is like that in Fig. 4(b), but the difference is that the output ports are on the right side of the structure. To unify the variables, the number of layers in the B domain of the output port is designed to be 1. We place red-marked point sources in the Fig. 5(d) to excite the TVWSs, which propagate to the output port along the top and the bottom waveguides, respectively. The number of layers in the B domain of the top waveguide is equal to 1 ($x$$_1$=1) and the number of layers in the bottom waveguide is equal to 2 ($x$$_2$=2). While maintaining the position of the bottom waveguide, we can obtain more splitting ratios by moving the position of the upper waveguide \cite{he2020topologically}. It should be noted that simply changing the relative positions of the upper and bottom waveguides does not result in more splitting ratios. Instead, by simultaneously adjusting the new adjustment DOF which includes the layer numbers of the upper and bottom waveguides and their relative positions, we can obtain the topological photonic power splitter with arbitrary power splitting ratio. Figure 5(e) shows the field distribution for the difference between the relative positions of the top and the bottom waveguides equal to 4$a$ ($a$ is the lattice constant). The field strength at port 1 is stronger than that at port 2. As shown in Fig. 5(f), the power splitting ratio of the device shows different power splitting cases as the top waveguide position is shifting. We define the shift distance as 0 for the case in Fig. 5 (d). As the top waveguide is moved along the port 1 direction with the lattice constant $a$ as the moving unit, the power splitting ratio of the device shows a nearly linear variation. As the number of layers of the bottom waveguide increases, the top waveguide moving range increases, and finer power splitting ratios can be obtained. Based on the results, the topological photonic power splitter with arbitrary power splitting ratios requires only a simple adjustment of the number of layers of the upper and bottom waveguides and their relative positions, which is important for the integrated photonic devices on chip.

\begin{figure}[h!]
	\centering\includegraphics[width=13cm]{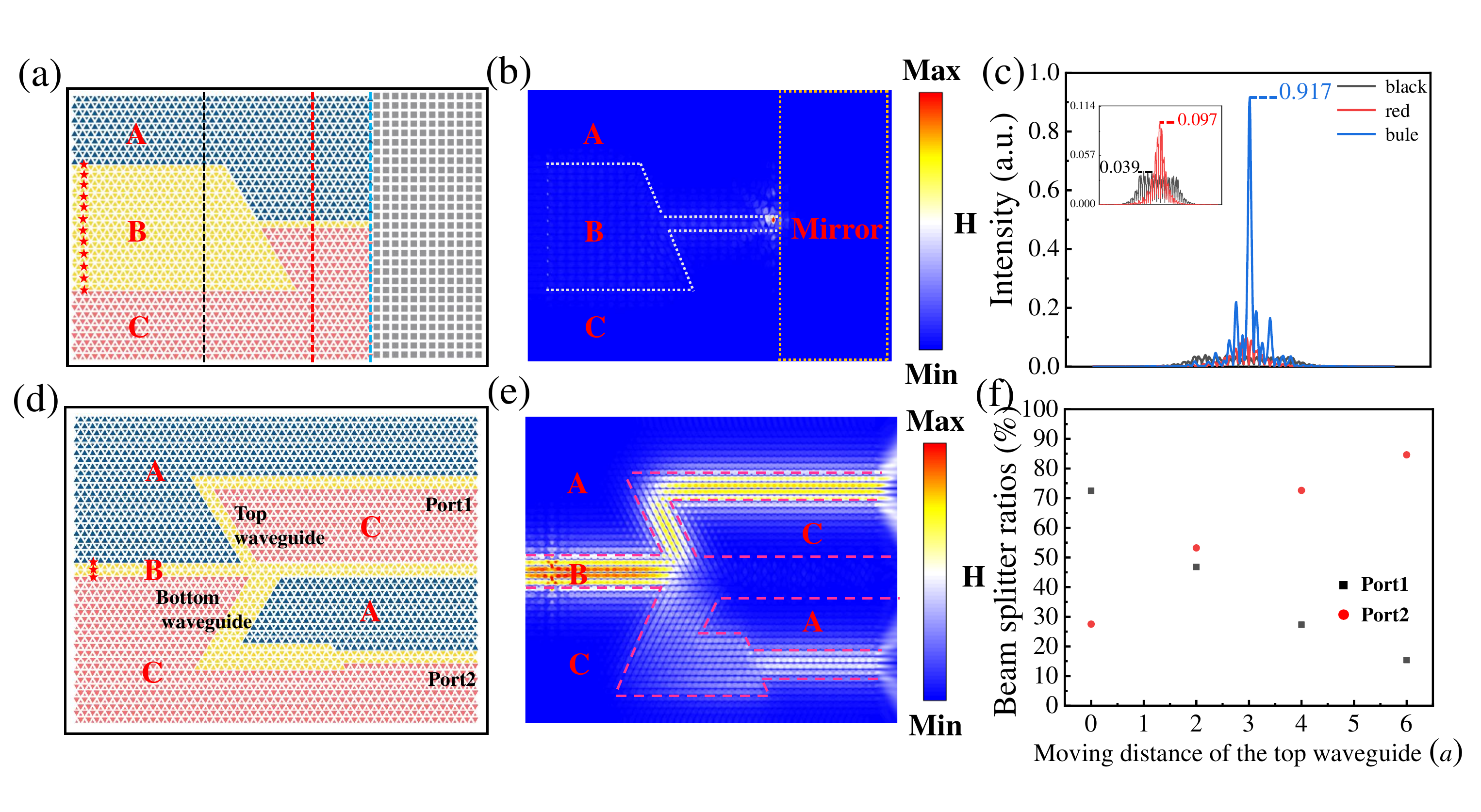}
	\caption{(a) Schematic of a topological concentrator. The width of B domain abruptly changes from 9 layers to 0.5 layer in the middle of the device. The waveguide with x=0.5 in the B domain is terminated by a mirror made up of a photonic crystal structure. Red stars: point sources. (b) Simulated H field distributions in the topological concentrator at 302.08 THz. (c) Simulated intensity profiles along black, red, and blue dashed lines in (a). (d) Schematic of a topological photonic power splitter. The width of B domain of top waveguide is 1 and that of bottom waveguide is 2. Red stars:  point sources. (e) Simulated H field distributions in the topological photonic power splitter at 302.08 THz. (f) The power splitting ratio of the device varies with the top waveguide shift in units of the lattice constant $a$.}
\end{figure}

Design parameters of all the devices mentioned above are experimentally feasible. In our work, the crystal constant in our design is 280 nm, the side lengths of the triangular airholes are 196 nm, 151.2 nm and 84 nm, and the thickness is 150 nm, which corresponds to a center operating frequency of 300 THz for the valley-locked waveguides. The devices with a thickness of 150 nm do not need a deep etching process, without the aspect ratio fabrication challenge compared to the devices for developing 6G technologies. \cite{tan2022interfacial,kumar2022phototunable,kumar2022topological,kumar2022active}. Photonic crystal waveguide structures with very similar design parameters have been fabricated and demonstrated in various applications, such as GaAs valley photonic crystal waveguide with light-emitting InAs quantum dots \cite{yamaguchi2019gaas}, protect transport at telecommunication wavelengths \cite{shalaev2019robust} and broadband Purcell enhancement \cite{xie2021topological}. Therefore, the device design with topologically protected waveguides is feasible and can be experimentally implemented.
\section{CONCLUSION}

In summary, we proposed a design on integrated optical devices by using topological valley-locked waveguide with the introduction of width DOF to manipulate the transport of light. Firstly we demonstrated that TVWSs possess gapless dispersions, momentum-valley locking and robustness. Then valley-locked energy concentrators and energy power splitters with arbitrary power splitting ratios have been designed on-chip with a higher flexibility by introducing width DOF, which provides opportunities for designing various kinds of devices for different functions. Comparing with previous topological waveguides, topologically protected photonic waveguides with the width DOF enable flexible design of on-chip integrated devices. More importantly, it can be flexibly interfaced with existing photonic devices, which promote the development of on-chip integrated optical networks.

\paragraph{Funding}
National Key Research and Development Program of China (Grant No. 2021YFA1400700); National Natural Science Foundation of China (Grants Nos. 62025507, 11934019, 92250301, 11721404, 62175254, 12174437, 12204020); Chinese Academy of Sciences (Grant No. XDB28000000); China Postdoctoral Science Foundation (Grant No. 2022M710234).
	
\paragraph{Disclosures}
The authors declare no conflicts of interest.

\paragraph{Data Availability}
Data underlying the results presented in this paper are not publicly available at this time but may
be obtained from the authors upon reasonable request.

\end{document}